\documentclass[12pt,a4paper]{article}

\usepackage{amssymb}
\usepackage{epsf}

\begin{document}

\title{Analytic approach to the motion of cosmological phase transition fronts}
\author{ {\large Ariel M\'egevand\thanks{%
Member of CONICET, Argentina. E-mail address: megevand@mdp.edu.ar}~ and
Alejandro D. S\'anchez\thanks{%
Member of CONICET, Argentina. E-mail address: sanchez@mdp.edu.ar}} \\
{\normalsize \emph{IFIMAR (CONICET-UNMdP)}, }\\
{\normalsize \emph{Departamento de F\'{\i}sica, Facultad de Ciencias Exactas
y Naturales, UNMdP,} }\\
{\normalsize \emph{De\'an Funes 3350, (7600) Mar del Plata, Argentina} }}
\date{}
\maketitle

\begin{abstract}
We consider the motion of planar phase-transition fronts in
first-order phase transitions of the Universe.  We find the steady
state wall velocity as a function of a friction coefficient and
thermodynamical parameters, taking into account the different
hydrodynamic modes of propagation. We obtain analytical
approximations for the velocity by using the thin wall approximation
and the bag equation of state. We compare our results to those of
numerical calculations and discuss the range of validity of the
approximations. We analyze the structure of the stationary solutions.
Multiple solutions may exist for a given set of parameters, even
after discarding non-physical ones. We discuss which of these will be
realized in the phase transition as the stationary wall velocity.
Finally, we discuss on the saturation of the friction at
ultra-relativistic velocities and the existence of runaway solutions.
\end{abstract}

\section{Introduction}

In a first-order cosmological phase transition, bubbles nucleate and
expand, converting the high-temperature phase into the
low-temperature one (see, e.g., \cite{gw81,ah92,m00}). As bubbles
expand, latent heat is released at their  boundaries. This energy
raises the temperature and causes bulk motions  of the plasma. The
perturbations caused in the cosmic fluid by the nucleation and
expansion of bubbles generate a departure from thermal equilibrium.
This may give rise to a number of cosmic relics, such as a baryon
number asymmetry \cite{ckn93}, baryon inhomogeneities  \cite{w84},
magnetic fields \cite{gr01}, topological defects \cite{vs94}, or
gravitational waves \cite{gw,lms12}.

In general, the system can be described by a relativistic fluid and a
scalar field $\phi$ at finite temperature $T$
\cite{ikkl94,kl95,kl96}. The latter may be a Higgs field and acts as
an order parameter. At high temperatures the free energy
$\mathcal{F}(\phi,T)$ has a minimum $\phi_+(T)$ (in general,
$\phi_+\equiv 0$) and at low temperatures a different minimum
$\phi_-(T)$. In a first-order phase transition, the two minima
coexist in certain range of temperatures, separated by a barrier. In
the high-$T$ phase, the free energy density is given by $
\mathcal{F}_{+}(T)= \mathcal{F}(\phi_+(T),T)$, whereas in the low-$T$
phase, it is given by $\mathcal{F}_{-}(T)= \mathcal{F}(\phi
_{-}(T),T)$. The critical temperature $T_c$ is that for which
$\mathcal{F}_{+}(T_{c})=\mathcal{F}_{-}(T_{c})$. The phase transition
occurs when the temperature of the Universe reaches $T_c$. At
$T=T_c$, though, the nucleation rate vanishes, and bubbles
effectively begin to nucleate at some temperature $T_n$ below $T_c$
\cite{ah92,c77}.

The nucleated bubbles expand due to the pressure difference between
the two phases. In general, the bubble walls reach a terminal
velocity $v_w$  due to the friction with the surrounding plasma.
Recently, the hydrodynamics of the moving walls has received much
attention (see, e.g., \cite{ms09,bm09,ekns10,lm11,kn11}) due to the
interest in performing thorough calculations of the wall velocity and
the energy injected into bulk motions of the fluid. These quantities
are relevant for the generation of baryons and gravitational waves.
In Ref. \cite{bm09}, the ultra-relativistic velocity regime was
considered, and it was shown that a state of continuous acceleration
of the bubble wall is possible. Such ``runaway'' solutions may play
an important role in the generation of gravitational waves.

A realistic evaluation of the cosmological consequences of a phase
transition requires considering the dynamics as completely as
possible. Following the development of a phase transition involves
the calculation of several temperature-dependent quantities such as,
e.g., the pressure of the two phases and the bubble nucleation rate.
During the phase transition the temperature varies in time and space
due to the adiabatic cooling of the Universe and the release of
latent heat. As a consequence, one has to deal with a set of
integro-differential equations. Some of the involved variables are
very sensitive to approximations (for instance, the nucleation rate).
In order to avoid large errors, it is convenient to resort to
nontrivial numerical calculations for these quantities. On the other
hand, finding analytical approximations for other variables (e.g.,
the bubble wall velocity) provides a way of reducing the computation
time. Widely used simplifications include the thin wall approximation
and the bag equation of state \cite{ikkl94,kl95,kl96,gkkm84}. Even
with these approximations, it is not always possible to obtain
analytical results.

A considerable simplification is achieved by considering planar
walls. Analytic results for the planar case were found recently in
Ref. \cite{ms09} for  the wall velocity and in Ref. \cite{lm11}  for
the energy injected into the fluid. It is important to note that
considering spherical bubbles is not necessarily a better
approximation than considering planar walls. Although the spherical
symmetry is a good approximation for the initial stages of bubble
growth, some cosmologically interesting outcomes of the phase
transition are produced when bubbles collide and lose the spherical
symmetry. Moreover, losing the spherical symmetry is a requirement,
e.g., for the generation of turbulence or of gravitational waves. As
an explicit example, the ``envelope approximation'' for the
generation of gravitational waves in bubble collisions neglects the
overlap regions of colliding bubbles and follows only the evolution
of the uncollided bubble walls. For such a calculation, the
approximation of treating the walls as planar is, in principle, as
good as considering spherical bubbles (but less complicated). In
general, one does not expect important differences  (see Ref.
\cite{lm11} for a comparison of  different wall geometries).

In this paper we investigate the propagation of a planar phase
transition front in the plasma. We aim at finding analytical
approximations for the stationary velocity of the bubble wall as a
function of the friction and the thermodynamical parameters. The
present work is a continuation of the investigations of Ref.
\cite{ms09}, where we considered planar walls propagating as weak
deflagrations or weak detonations. Here we include into consideration
the case of supersonic Jouguet deflagrations \cite{kl95} and the
possibility that the walls run away (see Ref. \cite{ekns10} for a
recent study for spherical-symmetry walls). We also discuss here
whether the different solutions are physical or not, and which of
them will be realized as final stationary states during the phase
transition. We discuss on the validity of the analytical
approximations. The approximations are better for weaker solutions
than for those close to the Jouguet points. For comparison, we
consider some cases previously studied with numerical calculations
\cite{ikkl94,kl95,kl96}.

The paper is organized as follows. In section \ref{hydro} we consider
the equations for the profiles of the fluid and the bubble wall,
including a phenomenological friction term. We study the thin wall
limit. In section \ref{bag} we use the bag equation of state to
obtain a set of analytic equations for the wall velocity. In sections
\ref{test} and \ref{result} we present our results for the stationary
motion and compare them with those of the numerical works of Refs.
\cite{ikkl94,kl95,kl96}. In section \ref{test} we discuss the range
of validity of the analytical approximations and in section
\ref{result} we analyze the dependence of the wall velocity on the
thermodynamic parameters and the friction. In section \ref{runaway}
we consider a different phenomenological friction term, which takes
into account the fact that, in some models, the friction force
approaches a constant in the ultra-relativistic regime. Finally, we
conclude in section \ref{conclu}.

\section{Hydrodynamics and microphysics} \label{hydro}

We shall consider the motion of bubble walls at a given temperature
$T_n<T_c$, i.e., we shall regard the nucleation temperature $T_n$ as
a free parameter. Therefore, we shall not be concerned here with the
calculation of the amount of supercooling. All the thermodynamic
quantities (energy density, pressure, temperature, etc.) are derived
from the free energy density. We have two phases, characterized by
the minima $\phi_{\pm}$. Thus, the equation of state (EOS) is
different in each phase. For instance, the energy density is given by
$\rho_{\pm} \left( T\right) =
\mathcal{F}_{\pm}(T)-T\mathcal{F}_{\pm}^{\prime }(T)$, where a prime
indicates derivative with respect to $T$. The pressure is given by
$p_{\pm}(T)=-\mathcal{F}_{\pm}(T)$. The enthalpy density is given by
$w_{\pm}=\rho_{\pm}+p_{\pm}$, and the entropy density by
$s_{\pm}=w_{\pm}/T$. The speed of sound is given by $ c_{\pm}^{2}(T)
=\partial p_{\pm}/\partial \rho_{\pm}=p_{\pm}^{\prime}(T)/\rho_{\pm}
'(T)$. The latent heat is defined as $ L\equiv\rho _{+}\left(
T_{c}\right) -\rho _{-}\left( T_{c}\right) $.

In the regions separating the two phases the scalar field $\phi$
varies from $\phi_+$ to $\phi_-$. These interfaces are the walls of
expanding bubbles, i.e., the phase-transition fronts. In general,
temperature gradients arise during the phase transition, and the
temperature varies beyond the bubble walls. Thus, the system is
characterized by the scalar field $\phi(\mathbf{x},t)$, the
temperature $T\left( \mathbf{x},t\right) $ of the plasma (which we
treat as a perfect relativistic fluid), and the fluid velocity
$v\left( \mathbf{x},t\right) $. These  variables are governed by the
equations
\begin{eqnarray}
\partial _{\mu }\left( -T\frac{\partial \mathcal{F}}{\partial T}u^{\mu
}u^{\nu }+g^{\mu \nu }\mathcal{F}\right) +\partial _{\mu }\partial ^{\mu
}\phi \partial ^{\nu }\phi &=&0,  \label{constmunu} \\
\partial _{\mu }\partial ^{\mu }\phi +\frac{\partial \mathcal{F}}{\partial
\phi }+\tilde{\eta}Tf\left( \phi /T\right) u^{\mu }\partial _{\mu }\phi &=&0,
\label{eqfield}
\end{eqnarray}%
where $u^{\mu }=(\gamma ,\gamma \mathbf{v})$ is the four-velocity
field of the fluid, with $\gamma =1/\sqrt{1-v^{2}}$. Equation
(\ref{constmunu}) follows from energy-momentum conservation,
$\partial _{\mu }T^{\mu \nu }=0$, whereas Eq. (\ref{eqfield}) is the
equation of motion for $\phi $, where we have introduced a general
phenomenological damping term proportional to $\partial _{\mu }\phi $
in order to account for the friction force acting on the scalar
field. In section \ref{runaway} we shall consider a modification of
this term which does not grow with $\gamma$ for large $\gamma$. The
function $f$ and the dimensionless friction parameter $ \tilde{\eta}$
can be derived by considering the microphysics in specific models
\cite{ms09,ms10,fricthermal}.

We are interested in the steady-state motion of bubble walls.
Therefore, we assume stationary profiles moving with the wall at
constant velocity. We shall consider planar-symmetry fronts moving in
the $x$ direction. In the rest frame of the front, all time
derivatives vanish and Eqs. (\ref{constmunu}-\ref{eqfield}) become
\begin{eqnarray}
&-T\frac{\partial \mathcal{F}}{\partial T}\gamma ^{2}v =\mathrm{constant},
\label{cont1} \\
&-T\frac{\partial \mathcal{F}}{\partial T}\gamma ^{2}v^{2}-\mathcal{F}+
\frac{1}{2}\left(\frac{d\phi}{dx}\right) ^{2} =\mathrm{constant},
\label{cont2} \\
&\frac{d^2\phi }{dx^2}-\frac{\partial \mathcal{F}}{\partial
\phi }-\tilde{\eta}Tf\left( \phi /T\right) \frac{d\phi}{dx}
\gamma v =0.  \label{contfric}
\end{eqnarray}%
For a given model the static, 1-dimensional Eqs.
(\ref{cont1}-\ref{contfric}) or the dynamic, 4-dimensional equations
(\ref{constmunu}-\ref{eqfield}), can be integrated numerically, e.g.,
using a lattice \cite{ikkl94,kl95,kl96}. However, it is very useful
to assume that the interface is infinitely thin, thus eliminating the
scalar field profile from hydrodynamics considerations. Assuming a
thin wall is in general a good approximation, as the wall width is
much smaller than the  width of the fluid profiles. Indeed, the
latter is determined by the dynamics of the phase transition and is
roughly given by the time scale $t$. In contrast, the width of the
wall, determined by the characteristic length of variation of $\phi
$, is roughly given by the scale $T^{-1}$. In general, $t$ is many
orders of magnitude larger than $T^{-1}$. As we shall see, taking the
thin wall limit in Eqs. (\ref{cont1}-\ref{cont2}) is trivial, whereas
doing so in Eq. (\ref{contfric}) requires additional approximations.

\subsection{Hydrodynamics}

\begin{figure}[htb]
\centering \epsfysize=8cm
\leavevmode \epsfbox{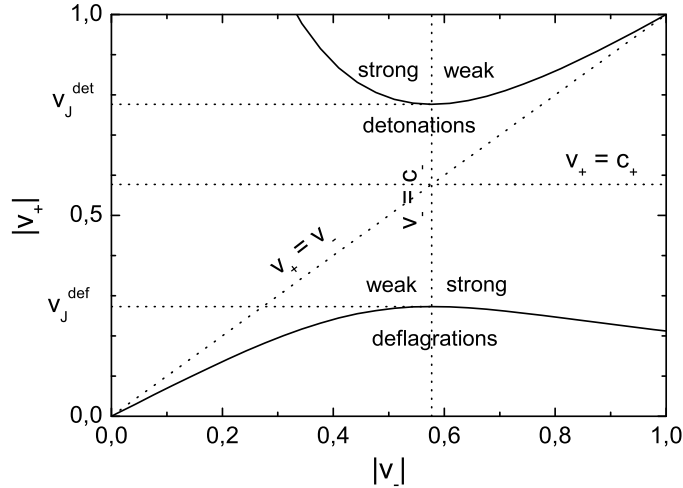}
\caption{$v_{+}$ vs $v_{-}$ for the bag EOS ($c_+=c_-=1/\sqrt{3}$),
for $\alpha_+=0.1$. The upper branch corresponds to
detonations and the lower to deflagrations.}
\label{figvmavme}
\end{figure}

Equations (\ref{cont1}) and (\ref{cont2}) relate the fluid variables
on each side of the wall (in the rest frame of the wall). We shall
use a $+$ sign for variables just in front of the wall and a $-$ sign
for variables just behind the wall. Since $\phi '$ vanishes outside
the interface, we have
\begin{eqnarray}
w_{-}v_{-}\gamma _{-}^{2} &=&w_{+}v_{+}\gamma _{+}^{2},  \label{disc1} \\
w_{-}v_{-}^{2}\gamma _{-}^{2}+p_{-} &=&w_{+}v_{+}^{2}\gamma _{+}^{2}+p_{+}.
\label{disc2}
\end{eqnarray}%
These equations give $v_{+}$ as a function of $v_{-}$. The solutions
have two branches (see Fig. \ref{figvmavme}), called
\emph{detonations }and \emph{deflagrations}. For detonations the
incoming flow is faster than the outgoing flow ($|v_{+}|>|v_{-}|$).
The value of $|v_{+}|$ is supersonic in all the range $ 0<|v_{-}|<1$,
and has a minimum at the \emph{Jouguet point} $|v_{-}|=c_{-}$. The
minimum value of $|v_{+}|$ is the Jouguet velocity
$v_{J}^{\mathrm{\det } }$, with $v_{J}^{\mathrm{\det }}>c_{+}$. For
deflagrations we have $ |v_{+}|<|v_{-}|$, and $|v_{+}|$ has a maximum
value $v_{J}^{\mathrm{def} }<c_{+}$ at the Jouguet point
$|v_{-}|=c_{-}$. The hydrodynamical process is called \emph{weak} if
the velocities $v_{+}$ and $v_{-}$ are either both supersonic or both
subsonic. Otherwise, the hydrodynamical process is called
\emph{strong}.

We may also have discontinuities in the fluid profiles out of the
bubble wall, which are called \emph{shock fronts}. In the reference
frame of these surfaces Eqs. (\ref{disc1}-\ref{disc2}) apply. The
functions $w(T)$ and $p(T)$ are the same on both sides of the surface
(since we have the same phase), but the temperature is discontinuous.

Away from the phase transition front, the field $\phi$ is constant
and Eq. (\ref{constmunu}) becomes
\begin{eqnarray}
\partial _{t}\left( w\gamma ^{2}-p\right) +\partial _{x}\left( w\gamma
^{2}v\right) &=&0,  \label{fluid1} \\
\partial _{t}\left( w\gamma ^{2}v\right) +\partial _{x}\left[ \left( w\gamma
^{2}v^{2}+p\right) \right] &=&0.  \label{fluid2}
\end{eqnarray}%
Since there is no characteristic distance scale in Eqs.
(\ref{fluid1}-\ref{fluid2}), it is usual to assume the
\emph{similarity condition} \cite{landau}, namely, that $w,p$ and $v$
depend only on $\xi =x/t$. Using the relation $p^{\prime }=c^{2}\rho'
$, where a prime indicates derivative with respect to $\xi $, we
obtain the equation for the velocity profile (see e.g. \cite{lm11})
\begin{equation}
\left[ \left( \frac{\xi -v}{1-\xi v}\right) ^{2}-c^{2}\right] v^{\prime
}=0.
\end{equation}%
Therefore we have either $v^{\prime }=0$, which gives constant solutions
\begin{equation}
v( \xi ) =\mathrm{constant} , \label{const}
\end{equation}%
or $( \xi -v) /( 1-\xi v) =\pm c$, which gives two additional
solutions. Of these, only the one corresponding to the $+$ sign will
be realized for the boundary conditions of our problem,
\begin{equation}
v_{\mathrm{rar}}( \xi ) =\frac{\xi -c}{1-\xi c},
\label{raref}
\end{equation}
which corresponds to a rarefaction profile. The speed of sound $c$ in
principle depends on $T$ and, thus, may depend on $\xi$. From Eqs.
(\ref{fluid1}-\ref{fluid2}) we also obtain the equation for the
enthalpy profile,
\begin{equation}
\frac{w^{\prime }}{w}=\left( \frac{1}{c^{2}}+1\right) \frac{\xi -v}{
1-\xi v}\gamma ^{2}v^{\prime },  \label{eqenth}
\end{equation}%
which can be readily integrated to obtain the profiles of the
thermodynamical variables. In particular, for $v=\mathrm{constant}$
we have $T=\mathrm{constant}$.

\begin{figure}[hbt]
\centering
\epsfxsize=16.5cm \leavevmode \epsfbox{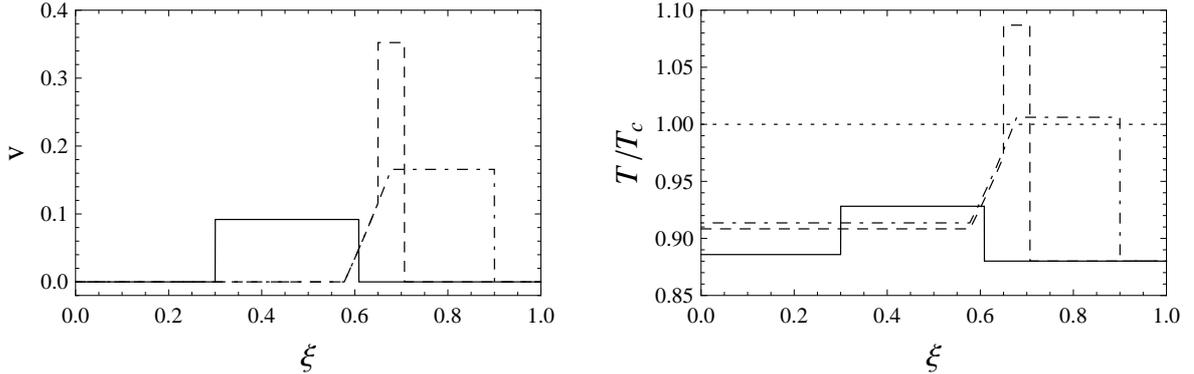}
\caption{Some fluid profiles for the bag EOS with  $\alpha
_{n}=0.1$ and $\alpha_c=0.06$ ($T_n/T_c\simeq 0.88$). Solid lines correspond to a
weak deflagration with $v_w=0.3$, dashed lines to
a Jouguet deflagration with $v_w=0.65$, and dashed-dotted lines to a weak detonation with
$v_{w}=0.9$. Left: the fluid velocity. Right: the fluid temperature.
The dotted line indicates the critical temperature.}
\label{profiles}
\end{figure}

The fluid velocity and temperature profiles are constructed with the
solutions (\ref{const}-\ref{eqenth}), using the matching conditions
(\ref{disc1}-\ref{disc2}) and appropriate boundary conditions. The
usual boundary conditions consist of a vanishing fluid velocity far
behind the moving wall (at the center of the bubble) and far in front
of the wall, where information on the bubble has not arrived yet. The
temperature far in front of the wall can be determined from the
dynamics of the phase transition. We shall assume it is given by the
nucleation temperature $T_{n}$ of the bubble. Three kinds of
solutions are compatible with all these requirements (see \cite{lm11}
for details):

\textbf{A detonation.} The wall is supersonic and the fluid in front
of it is unperturbed (see Fig. \ref{profiles}). Therefore, the fluid
velocity in front of the wall vanishes, and we have $v_{w}=-v_+$,
with $|v_{+}|\geq v_{J}^{\det }$. It turns out that the fluid profile
is only compatible with a weak detonation. Thus, the outgoing flow is
also supersonic, $|v_{-}|\geq c_{-}$. Behind the wall, the fluid
velocity is a constant between a certain point $\xi _{0}$ and  $\xi
_{w}= v_{w} $. At $\xi =\xi _{0}$ the fluid velocity matches the
rarefaction solution (\ref{raref}), which vanishes at $\xi =c_{-}$.
For $\xi<c_-$ we have $v=0 $. As we shall see, the fluid profile
behind the wall does not play a role in the determination of the
detonation wall velocity.

\textbf{A ``traditional'' deflagration.} For this solution the fluid
behind the wall is at rest, so $v_{w}=-v_{-}$. The deflagration could
in principle be weak,  Jouguet, or strong (however, the latter seems
to be unstable \cite{ikkl94,kl95,kl96}). The fluid velocity in front
of the wall is a constant up to a shock front $\xi_{\mathrm{sh}}$,
which moves supersonically. Beyond the shock, the fluid is still
unperturbed.

\textbf{A supersonic deflagration.} The hydrodynamical solution is a
Jouguet deflagration, i.e., $|v_{+}|<|v_{-}|$ and $|v_{-}|=c_{-}$.
Hence, we have $ |v_{+}|=v_{J}^{\mathrm{def}}<c_{+}$. In this case,
neither the fluid behind the wall nor in front of it is at rest, and
the Jouguet condition $v_-=-c_-$ replaces the condition
$v_{-}=-v_{w}$ of the traditional deflagration. Since the wall moves
at the speed of sound with respect to the fluid behind it, and the
fluid also moves (dragged by the wall), the wall velocity is always
supersonic. The fluid profile behind the wall is given by the
rarefaction solution (\ref{raref}) between $c_-$ and $ \xi _w$. In
front of the wall, the fluid velocity is a constant  up to the shock
front.

Fixing the thermodynamical parameters, the latter solution exists for
$c_-\leq v_w\leq v_J^{\mathrm{det}}$. As the wall velocity of the
supersonic deflagration approaches the limit $\xi _{w}=c_{-}$, the
rarefaction wave disappears and this solution matches the traditional
deflagration. As $\xi _{w}$ increases, the shock front and the
phase-transition front become closer. In the limit $\xi
_{w}=v_{J}^{\det }$, the shock wave disappears and the solution
matches the detonation.

\subsection{Microphysics} \label{micro}

In order to obtain analytical results for the wall velocity, we shall
implement the thin wall approximation in the friction equation
(\ref{contfric}). Following Ref. \cite{ms09}, we multiply Eq.
(\ref{contfric}) by $d\phi/dx$, integrate across the wall, and use
$({\partial \mathcal{F}}/{\partial \phi }) ({d\phi }/{dx})=
{d\mathcal{F}}/{dx}-({\partial \mathcal{F}}/{\partial T})({dT}/{dx})$
\cite{ikkl94}. We obtain
\begin{equation}
p_{+}-p_{-}-\int \left( -\frac{\partial \mathcal{F}}{\partial T}\right)
dT-\int \tilde{\eta}Tf(\phi /T)\left( \phi ^{\prime }\right) ^{2}v\gamma
dx=0, \label{fricint}
\end{equation}%
where $\phi'\equiv d\phi/dx$. We will make approximations for the
integrals in Eq. (\ref{fricint}), so that we can express the result
in terms of the values of the variables outside the wall.

First, we write the friction term as
\begin{equation}
-\int \tilde{\eta}Tf(\phi /T)\left( \phi ^{\prime }\right) ^{2}\, v\gamma\,
dx\equiv \eta \frac{|v_{+}|\gamma _{+}+|v_{-}|\gamma _{-}}{2}
\label{aproxvmedia}
\end{equation}%
(notice that the fluid velocity $v$ in the rest frame of the wall is
negative). The friction coefficient $\eta $ is usually approximated
by its value in the limit $T_{+}=T_{-}=T_{n}$, i.e., when  the
effects of hydrodynamics are neglected and only microphysics is
considered. This is the limit in which the fluid is left unperturbed
by the wall, and is valid for very weak solutions, for which
$v_{+}=v_{-}=-v_{w}$. According to Eqs.
(\ref{fricint}-\ref{aproxvmedia}), the wall velocity is given in this
limit by
\begin{equation}
v_{w}\gamma _{w}|_{\mathrm{micro}}=\frac{ p_{-}(T_{n})-p_{+}(T_{n})}{\eta},
\end{equation}%
with $\eta$ given by
\begin{equation}
\eta =\tilde{\eta}T\int f(\phi /T)\left( \phi ^{\prime }\right) ^{2}dx.
\label{etamicro}
\end{equation}
We shall consider $\eta$ as a  constant  in Eq. (\ref{aproxvmedia}),
which can be calculated for a specific model using the approximation
(\ref{etamicro}). Similarly, for the first integral in Eq.
(\ref{fricint}) we shall use a linear approximation for the entropy
density inside the wall, which gives
\begin{equation}
\int \left( -\frac{\partial \mathcal{F}}{\partial T}\right) dT\simeq
\frac{\left( s_{+}+s_{-}\right) \left( T_{+}-T_{-}\right) }{2}.
\label{intentr}
\end{equation}

With these approximations, Eq. (\ref{fricint}) becomes \cite{ms09}
\begin{equation}
p_{+}-p_{-}-\frac{s_{+}+s_{-}}{2}\left( T_{+}-T_{-}\right) +\frac{1}{2}\eta
\left( |v_{+}|\gamma _{+}+|v_{-}|\gamma _{-}\right) =0.  \label{fric}
\end{equation}
This expression is valid for any specific choice of the effective
potential, since the information on the EOS is encoded in the
variables $p$ and $s$. Furthermore, considering a different
phenomenological damping only amounts to modifying the function of
$v_+$ and $v_-$ in the last term and is straightforward, as we shall
see in section \ref{runaway} with an example. Therefore, Eq.
(\ref{fric}) is readily applicable to a large variety of models.

The approximation (\ref{etamicro}) neglects the variation of the
fluid velocity inside the wall, since it is obtained from Eq.
(\ref{aproxvmedia}) by replacing $v\gamma$ with its mean value. We
expect this approximation to be better for weak solutions. Jouguet
and strong processes cause larger perturbations of the fluid, and
will give larger deviations. Indeed, as can be seen in Fig.
\ref{figvmavme}, for weak solutions the difference between $v_+$ and
$v_-$ is maximal at the Jouguet point. Similarly, Eq. (\ref{intentr})
approximates the entropy density inside the wall by the mean value
$(s_++s_-)/2$. We expect this approximation to be good for weaker
solutions and to fail perhaps for Jouguet or stronger solutions, for
which the difference between $+$ and $-$ variables can be large. As
we shall see, since only weak or, at most, Jouguet processes are
actually realized in the phase transition, the strongest deviations
from the correct results will indeed occur for velocities which are
close to the Jouguet deflagration or detonation.

The friction force was calculated for some specific models for the
case of the electroweak phase transition \cite{fricthermal}. In the
thin wall approximation, the wall profile can be estimated by
neglecting the temperature variation and the last term in Eq.
(\ref{contfric}). Thus, we have $d\phi /dx=-\sqrt{
2\Delta\mathcal{F}( \phi ,T) }$, with
\begin{equation}
\Delta\mathcal{F}( \phi,T )\equiv \mathcal{F}(\phi ,T)-\mathcal{F}(0,T),
\end{equation}
and Eq. (\ref{etamicro}) gives
\begin{equation}
\eta \simeq \tilde{\eta}T\int f(\phi /T)\sqrt{2\Delta\mathcal{F}( \phi,T )}
d\phi .  \label{eta}
\end{equation}%
Roughly, we have $\eta \approx \tilde{ \eta}T\sigma $, where $\sigma
=\int \phi ^{\prime 2}dx$ is the surface tension of the bubble wall.
This corresponds to setting $f(\phi /T)= 1$. The function $f$,
though, can make a quantitative difference (see, e.g.,
\cite{ms09,ms10}). However, since we are not going to consider any
particular model but rather take $\tilde{\eta}$ as a free parameter,
a different  $f$ in Eq. (\ref{eta}) only amounts to a redefinition of
$\tilde{\eta}$.

\section{The bag EOS} \label{bag}

Equations (\ref{disc1}), (\ref{disc2}) and (\ref{fric}) can be solved
for the wall velocity once the equation of state of the system is
known. It is convenient to use as an approximation the bag EOS,
\begin{equation}
\mathcal{F} _{+}\left( T\right) =-a_{+}T^{4}/3+\varepsilon, \ \
\mathcal{F}_{-}\left( T\right) =-a_{-}T^{4}/3.  \label{eos}
\end{equation}
This simplification allows to find analytical expressions for the
solutions. In this model the latent heat is given by $L=4\varepsilon
$ and the speed of sound is a constant $c=1/\sqrt{3}$ in both phases.
It is customary to express the results as functions of the variable
$\alpha \equiv \varepsilon /\left( a_{+}T^{4}\right) $. As discussed
in Refs. \cite{ms09,ms10}, for applications it is convenient to use
$L$ instead of $\varepsilon $.
Therefore, we define the parameters%
\begin{equation}
\alpha _{c}=\frac{L}{4a_{+}T_{c}^{4}},\quad \alpha _{+}=
\frac{L}{4a_{+}T_{+}^{4}},\quad \alpha _{n}=\frac{L}{4a_{+}T_{n}^{4}}.
\end{equation}
For the bag EOS Eqs. (\ref{disc1}) and (\ref{disc2}) give
\begin{equation}
v_{+}=\frac{\frac{1}{6v_{-}}+\frac{v_{-}}{2}\pm \sqrt{\left( \frac{1}{6v_{-}}%
+\frac{v_{-}}{2}\right) ^{2}+\alpha _{+}^{2}+\frac{2}{3}\alpha _{+}-\frac{1}{%
3}}}{1+\alpha _{+}} , \label{vmavme}
\end{equation}%
which is plotted in Fig. \ref{figvmavme} for $\alpha_+=0.1$. The plus
sign corresponds to detonations and the minus sign to deflagrations.
The friction equation (\ref{fric}) can also be expressed in terms of
$v_{+} $, $v_{-}$, and $\alpha _{+}$,
\begin{equation}
\frac{4v_{+}v_{-}\alpha _{+}}{1-3v_{+}v_{-}}-\frac{2}{3}\left( 1+\frac{s_{-}%
}{s_{+}}\right) \left( 1-\frac{T_{-}}{T_{+}}\right) +\frac{2\alpha _{+}\eta
}{L}\left( \left\vert v_{+}\right\vert \gamma _{+}+\left\vert
v_{-}\right\vert \gamma _{-}\right) =0,  \label{fricbag}
\end{equation}%
with
\begin{equation}
\frac{s_{-}}{s_{+}}=\frac{a_{-}}{a_{+}}\left( \frac{T_{-}}{T_{+}}\right)
^{3}\quad \mathrm{and}\quad \frac{T_{-}}{T_{+}}=\left[ \frac{a_{+}}{a_{-}}%
\left( 1-\alpha _{+}\frac{1+v_{+}v_{-}}{1/3-v_{+}v_{-}}\right) \right]
^{1/4}.  \label{tmatme}
\end{equation}%
The latter expressions depend on the ratio $a_-/a_+$, which
introduces a dependence on the parameter $\alpha_c$,
\begin{equation}
a_-/a_+=1-3\alpha_c.
\end{equation}

From Eqs. (\ref{vmavme}-\ref{fricbag}) we can find the velocities
$v_{+}$ and $v_{-}$ as functions of $\eta /L$, $\alpha _{+}$ and
$\alpha_c$. However, the variable $\alpha_+$ can be eliminated, since
the temperature $T_+$ in front of the wall is a function of the
nucleation temperature $T_n$. The relation between $\alpha _{+}$ and
$ \alpha _{n}$ depends on the type of hydrodynamic solution. For
detonations the temperature $T_{+}$ is just given by $T_+=T_n$, and
hence $\alpha _{+}=\alpha _{n}$. For deflagrations, $T_{+}$ is
related to $ T_{n}$ through the matching conditions at the shock
discontinuity,
\begin{equation}
v_{1}v_{2}=\frac{1}{3},\quad \frac{v_{1}}{v_{2}}=
\frac{3T_{n}^{4}+T_{+}^{4}}{3T_{+}^{4}+T_{n}^{4}},
\end{equation}%
where $v_{1}$ is the velocity of the outgoing flow in the reference
frame of the shock, and $v_{2}$ that of the incoming flow. In the
rest frame of the bubble center, the fluid velocity in front of the
shock vanishes (see Fig. \ref{profiles}). Hence, the  velocity of the
shock is given by $v_{\mathrm{sh}}=-v_2$. In the shock-wave region
the fluid velocity is a constant and, thus, can be obtained either
from the velocity $v_+$ or from the velocity $v_1$. This gives the
equation
\begin{equation}
\frac{v_{w}-|v_{+}|}{1-|v_{+}|v_{w}}=\frac{\sqrt{3}\left( \alpha _{n}-\alpha
_{+}\right) }{\sqrt{\left( 3\alpha _{n}+\alpha _{+}\right) \left( 3\alpha
_{+}+\alpha _{n}\right) }}.  \label{vfl}
\end{equation}%

Fixing the friction and the critical temperature, we can use the
above equations to obtain $v_w$ as a function of $\alpha_n$ as
follows. For detonations we have $v_+=-v_w$. We can eliminate $v_-$
from Eq. (\ref{vmavme}), and then obtain $v_w$ from Eq.
(\ref{fricbag}) as a function of $\alpha_+=\alpha_n$. For traditional
deflagrations, we have $v_{-}=-v_{w}$, so we can eliminate $v_+$
using Eq. (\ref{vmavme}) and obtain $v_w$ from Eq. (\ref{fricbag}) as
a function of $\alpha_+$. Then, Eq. (\ref{vfl}), together with Eq.
(\ref{vmavme}), can be used to obtain $\alpha _{+}$ as a function of
$\alpha _{n}$ and $v_w$. For Jouguet deflagrations, we have
$v_{-}=-1/\sqrt{3}$ fixed, so Eq. (\ref{vmavme}) alone gives $v_{+}$
as a function of $\alpha _{+}$, i.e.,
$v_{+}=v_{J}^{\mathrm{def}}\left( \alpha _{+}\right) $. Therefore,
Eq. (\ref{fricbag}) gives already the value of $\alpha _{+}$ (as a
function of the parameters $\eta$ and $\alpha_c$). In this case, Eq.
(\ref{vfl}) can be used to obtain the wall velocity as a function of
$\alpha _{+}$ and $\alpha _{n}$.

\section{General structure of the solutions} \label{test}

Solving Eqs. (\ref{vmavme}-\ref{vfl}) just amounts to finding the
roots of algebraic equations, thus avoiding time-consuming numerical
calculations. This is valuable when considering the development of a
phase transition. However, in the way from Eqs.
(\ref{cont1}-\ref{contfric}) to Eqs. (\ref{vmavme}-\ref{vfl}) we have
made several approximations. Therefore, it is important to determine
the range of validity of these results.

\begin{figure}[hbt]
\centering
\epsfxsize=11cm \leavevmode \epsfbox{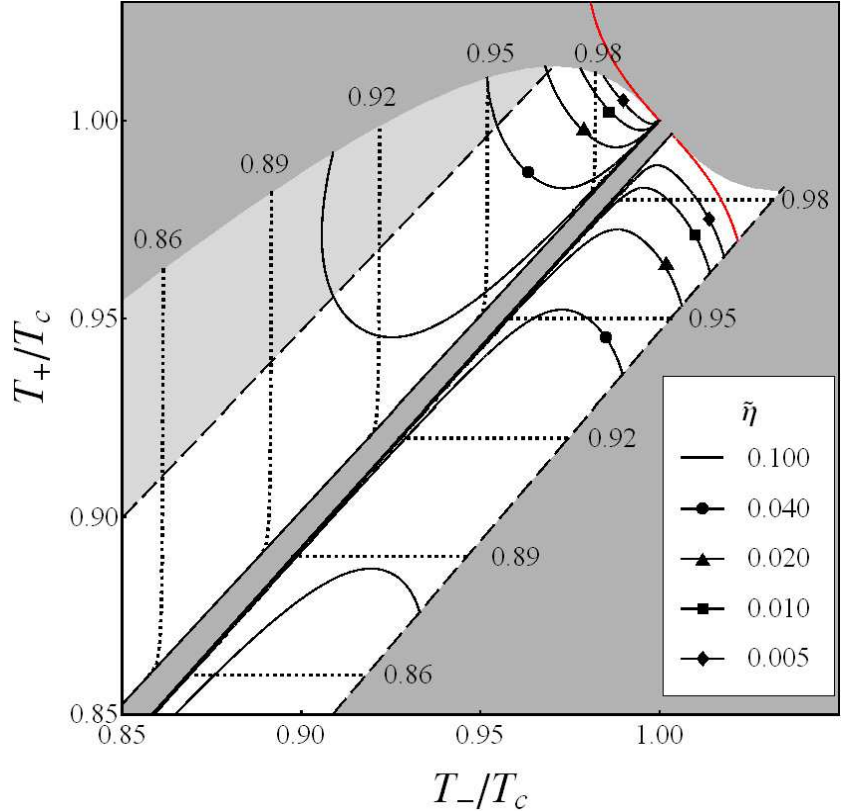}
\caption{The solutions in the $T_{+}T_{-}$-plane for $a_{+}=(\pi
^{2}/90)g_{\ast }$ with $g_{\ast }=51.25$, $L=0.1T_{c}^{4}$, and
$\eta=\tilde{\eta}T\sigma$, with $\sigma
=0.1T_{c}^{3}$. The region with the
dark shade is forbidden by kinematics
or by the non-negativity of entropy production.
The upper allowed region corresponds to deflagrations, and the
lower one to detonations. The Jouguet processes are indicated by dashed
lines, and the region with a lighter shade corresponds to strong
deflagrations. Dotted lines correspond to fixed values of
$T_{n}/T_{c}=0.86,0.89,0.92,0.95$ and 0.98 as indicated, and
solid lines to $\tilde{\protect\eta}$ fixed (see explanation in the text).
The red line corresponds to $\tilde{\eta}=0$.}
\label{figintersec}
\end{figure}

It is useful to consider the structure of the stationary states in
the $T_{+}T_{-}$-plane. Following \cite{kl95}, we consider, on the
one hand, the solutions which satisfy the energy-momentum
conservation and the friction equation, i.e., Eqs.
(\ref{cont1}-\ref{contfric}), but for which the boundary conditions
have not been imposed. For the bag EOS and our analytical
approximations, this amounts to considering Eq. (\ref{fricbag}), with
$v_+$ and $v_-$ expressed in terms of $T_+$ and $T_-$,
\begin{eqnarray}
v_-^2=\frac{(1/3 - \alpha_+ - r/3) (1 + \alpha_+ + r/3)}{(1 +
\alpha_+ - r) (r + 1/3 - \alpha_+)} \label{vme2} \\
v_+^2=\frac{(1/3 - \alpha_+ - r/3) (r - \alpha_+ + 1/3)}{(1 +
\alpha_+ - r) (1 + r/3 + \alpha_+)}, \label{vma2}
\end{eqnarray}
with
\begin{equation}
r\equiv\frac{a_-}{a_+} \frac{T_-^4}{T_+^4}. \label{r}
\end{equation}
This gives curves of $\eta=\mathrm{constant}$ in the
$T_{+}T_{-}$-plane (see Fig. \ref{figintersec}). On the other hand,
we consider the solutions  which satisfy the energy-momentum
conservation equations, Eqs. (\ref{cont1}-\ref{cont2}), and the
boundary conditions, but for which the friction equation
(\ref{contfric}) has not been imposed. This gives curves of fixed
$T_n$ in the $T_{+}T_{-}$-plane. For detonations, these are curves of
constant $T_{+}=T_{n}$. For traditional deflagrations, the curves are
given by Eq. (\ref{vfl}) with $v_w=|v_-|$, and with $v_{\pm}$ given
by Eqs. (\ref{vme2}-\ref{r}). Jouguet solutions are given by the
condition $v_-^2=1/3$. The values of $T_{+}$ and $T_{-}$ are
constrained by the conditions $0<v_{\pm}^2<1$ and by the production
of entropy at the phase transition front, $s_-|v_-|\gamma_-\geq
s_+|v_+|\gamma_+$.

The result is shown in Fig. \ref{figintersec}. The dark grey region
is forbidden by kinematics and by the condition of non-negative
entropy production. There are two allowed regions. The upper one
corresponds to deflagrations and the lower one to detonations. Weak
deflagrations (white zone) are separated from strong deflagrations
(light grey zone) by a dashed line which indicates Jouguet processes.
The deflagration region is delimited by the line of zero entropy
production (curved part of the boundary) and by the line of $v_-=0$
(straight part of the boundary). The lower white region,
corresponding to weak detonations, is delimited by the Jouguet
condition (dashed line), the zero entropy-production condition
(curved boundary), and the condition $v_+=1$ (upper straight
boundary). Solid lines represent the solutions of constant $\eta$.
Dotted lines represent the solutions for a given temperature $T_{n}$
for traditional deflagrations and for detonations.
For detonations, the
dotted curves are horizontal lines due to the boundary condition
$T_{+}=T_{n}$. For the traditional deflagrations, the curves are
almost vertical lines, indicating that, although the temperature
$T_{+}$ can be quite higher than $T_{n}$ due to reheating in front of
the wall, the temperature $T_{-}$ inside the bubble is very close to
$T_{n}$ \cite{kl95}. If we plotted the dotted curves also for
supersonic Jouguet deflagrations, they would lie on the dashed line
separating weak and strong deflagrations.

The possible stationary states correspond to the intersections of
solid and dotted curves in Fig. \ref{figintersec}. Thus, we see that
for some values of the parameters (e.g., for $T_n/T_c=0.95$,
$\tilde{\eta}=0.04$) we will have multiple solutions for the wall
velocity \cite{ikkl94,kl95,kl96}. Notice also that solid lines in the
deflagration region approach asymptotically the kinematic boundary
corresponding to $v_-=0$. As a consequence, for large enough
$\tilde{\eta}$ or for $T_n$ close enough to $T_c$, there will always
be weak deflagration solutions. Similarly, solid lines in the
detonation region approach asymptotically the boundary of $v_+=1$.
Therefore, there will always be detonations for small enough friction
or strong enough supercooling. For intermediate values of
$\tilde{\eta}$ and $T_n/T_c$, it may happen that neither weak
deflagrations nor detonations exist, as e.g. for the case
$\tilde{\eta}=0.1$, $T_n/T_c=0.89$ (whereas for $T_n/T_c=0.86$ we
have a detonation and for $T_n/T_c=0.92$ we have a weak
deflagration). Generally, in such a case there will exist supersonic
Jouguet deflagrations.

In Refs. \cite{ikkl94,kl95,kl96}, the differential equations for
$\phi (x)$, $T(x)$, and $v(x)$ were solved numerically using a grid.
Furthermore, a $\phi$-dependent, quartic potential was considered as
an approximation for the free energy. Comparing the results of those
numerical computations with our results, we can test to what extent
the use of the bag EOS, the thin wall limit and the approximations
(\ref{aproxvmedia}-\ref{fric}) are valid. Thus, in Fig.
\ref{figintersec} we have considered ``QCD-type'' parameters used in
Refs. \cite{ikkl94,kl95,kl96} and we have set $f( \phi /T) =1$ in Eq.
(\ref{eta}), which gives a friction of the form $\eta
=\tilde{\eta}T\sigma $. The value of $a_{+}$ is given by $a_{+}=(\pi
^{2}/90)g_{\ast }$, where the number of effective degrees of freedom
in the high-temperature phase is $g_{\ast }=51.25$. The latent heat
is given by $L=0.1T_{c}^{4}$, and the surface tension is $\sigma
=0.1T_{c}^{3}$. The set of values for the friction and for the
nucleation temperature are the same as in Fig. 1 of Ref. \cite{kl95}.

The dotted lines, as well as the limits of the allowed region, are
qualitatively identical and quantitatively quite close to those of
Ref. \cite{kl95}. This indicates that the bag model is a good
approximation for the EOS, at least for QCD-type parameters. In the
white region (i.e., that of weak solutions) the solid curves are
quantitatively similar to those of Ref. \cite{kl95}. Qualitatively,
the curves deviate from those of Ref. \cite{kl95} as they approach
the dashed lines (i.e., as the solutions approach a Jouguet process).
This is most apparent for $\tilde{\eta}\rightarrow 0$. In the plot of
Ref. \cite{kl95}, for small friction the solid curves are parallel to
the border of the allowed region (in the top-right corner of our Fig.
\ref{figintersec}). This is because the entropy production is related
to gradients of $ \phi $ through the friction parameter
$\tilde{\eta}$. Indeed, combining Eqs. (\ref{cont1}-\ref{contfric}),
one obtains
\begin{equation}
T\frac{d}{dx}\left( -\frac{\partial \mathcal{F}}{\partial T}\gamma v\right) =
\tilde{\eta}T\left( \gamma v\right) ^{2}f\left( \phi /T\right) \phi ^{\prime
}\left( x\right) ^{2},  \label{entr}
\end{equation}%
which yields%
\begin{equation}
s_{-}|v_{-}|\gamma _{-}-s_{+}|v_{+}|\gamma _{+}=\tilde{\eta}\int \left( \gamma
v\right) ^{2}f\left( \phi /T\right) \phi ^{\prime }\left( x\right) ^{2}dx.
\end{equation}%
As a consequence, the solid curve for $\tilde{\eta}=0$ should
coincide with the limit between the white and grey regions,
corresponding to zero entropy production. Instead of that, in our
case this curve (red line) enters the allowed region for detonations
and enters the forbidden region for deflagrations. This is an
indication of the break-down of the approximation for the integral of
the entropy density across the wall, Eq. (\ref{intentr}), as weak
solutions approach a Jouguet process (see the discussion in
subsection \ref{micro}). However, as we shall see in the next
section, in most cases our analytical approximations give values of
the wall velocity which are qualitatively and quantitatively good,
even for solutions near the Jouguet point.

In Ref. \cite{kl95}, the solid lines in the $T_{+}T_{-}$-plane do not
penetrate the strong deflagration region but stop at the Jouguet
line. This means that the numerical code did not find any strong
deflagrations. As explained in Ref. \cite{kl95}, this is due to the
fact that no matter how well one tries to guess the correct solution,
the guess does not relax to a strong deflagration but rather changes
considerably to form a different type of solution. This is an
indication that strong deflagrations are unstable. In Fig.
\ref{figintersec} it is seen that we find strong deflagration
solutions. We believe this is not  a shortcoming of our
approximations, but rather due to the fact that analytical equations
allow to find any solution, even the unstable ones. In this region,
though, the departure of our solid lines from the actual curves is
probably large. In any case, being unstable, strong deflagrations are
of little interest.

\section{The  wall velocity} \label{result}

\begin{figure}[hbt]
\centering
\epsfysize=8cm \leavevmode \epsfbox{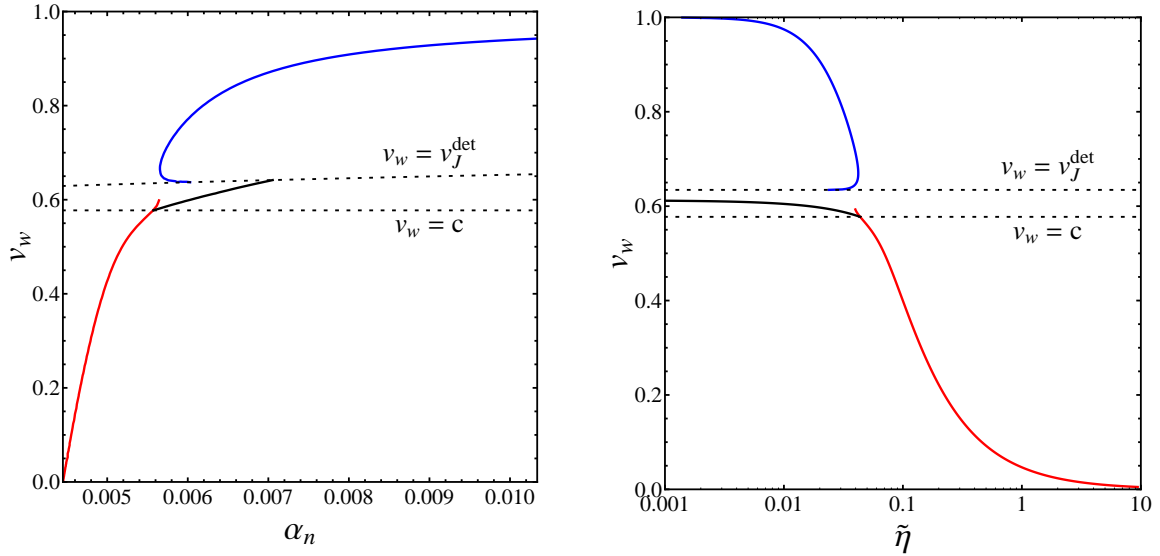}
\caption{The wall velocity for $g_{\ast }=51.25$, $L=0.1T_{c}^{4}$, and $
\protect\eta =\tilde{\protect\eta}T\protect\sigma $, with $\protect\sigma
=0.1T_{c}^{3} $. Detonations are plotted in blue, Jouguet deflagrations are
plotted in black, and traditional deflagrations are in red. The dotted
lines correspond to the sound and Jouguet velocities. Left: the wall
velocity as a function of $\protect\alpha _{n}$, for $\tilde{\protect\eta}
=0.05$. The range of $\protect\alpha _{n} $ correspond to values of the
temperature between $T_{n}=T_{c}$ and $T_{n}\simeq 0.8T_{c}$. Thus, the
$\alpha_n$-axis begins at the value $\alpha_n=\alpha _{c}\simeq 4.45\times
10^{-3}$. Right: the wall velocity as a function of the friction,
for $T_{n}=0.95T_{c}$ ($\protect\alpha _{n}\simeq 5.46\times 10^{-3}$).}
\label{figdetodefla}
\end{figure}

Let us consider now the wall velocity as a function of the
parameters. As can be seen in Fig. \ref{figdetodefla}, weak
traditional deflagrations (red curves) always exist for large
friction ($\tilde{\eta}\gtrsim 1$) or little supercooling
($\alpha_n\approx\alpha_c$). For lower values of the friction or
larger amounts of supercooling, the traditional deflagrations surpass
the speed of sound, becoming strong deflagrations. Shortly after
crossing the sound barrier, though, the red curves end due to the
condition of non-negative entropy production. At this point we always
have supersonic Jouguet deflagrations (black lines), which match the
traditional deflagrations at $v_w=c$. For small enough friction or
strong enough supercooling we always have detonations (blue lines).

\begin{figure}[hbt]
\centering
\epsfysize=8cm \leavevmode \epsfbox{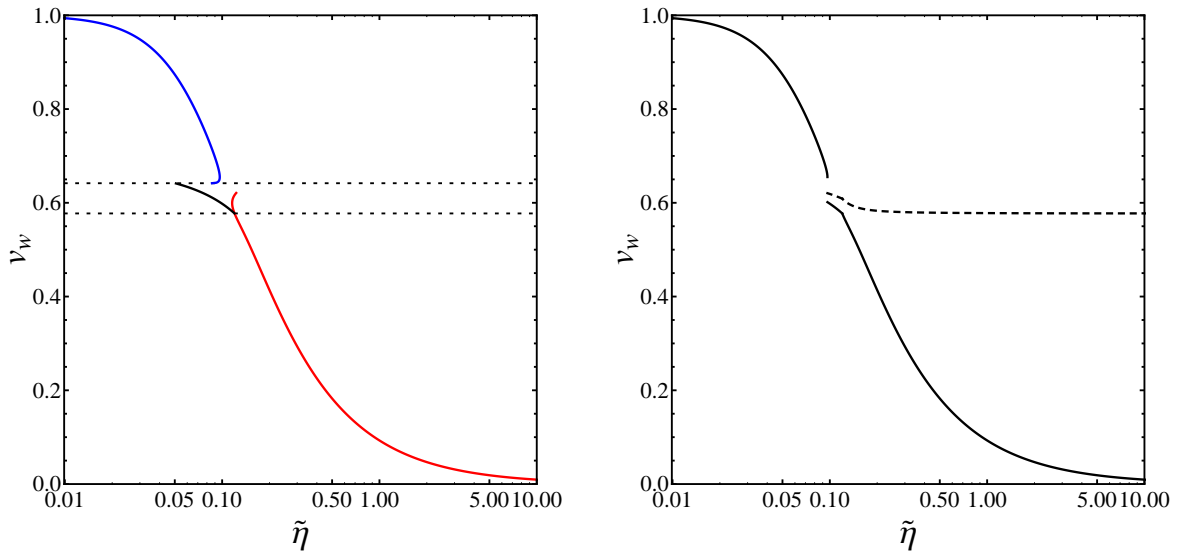}
\caption{The same as in Fig. \protect\ref{figdetodefla} (right), but for
$T_{n}=0.891T_{c}$. In the right panel only one of the multiple solutions
has been chosen. The dashed line corresponds to the velocity of the shock front.}
\label{figetaikkl}
\end{figure}

Notice that, for some values of $\eta $ and $\alpha_{n}$, there are
more than one stationary state.  This can be seen already in Fig.
\ref{figintersec}, since the solid and dotted curves intersect at
several points. In Fig. \ref{figdetodefla} we show all the solutions
for a given friction and supercooling. In some ranges of the
parameters we have, for instance, a deflagration and a detonation, or
a deflagration and two detonations. This is in agreement with the
results of Refs. \cite{ikkl94,kl95,kl96}. We may also have two
traditional deflagration solutions for a given set of parameters, as
shown in the left panel of Fig. \ref{figetaikkl}. One of these
solutions is a strong deflagration.

One could expect that it will be always possible to choose the
solution in such a way that, as the parameters are varied, one can go
continuously from a subsonic, weak deflagration to a supersonic,
Jouguet deflagration and then to a weak detonation. As seen in Figs.
\ref{figdetodefla} and \ref{figetaikkl}, one can indeed change
continuously from a weak deflagration to a supersonic Jouguet
deflagration. However, the Jouguet deflagration velocity does not
match the detonation velocity. The continuity of the deflagration
solutions is a consequence of the continuity of the profiles.
Consider the supersonic Jouguet deflagration. As the velocity
approaches the speed of sound $c$, the rarefaction wave vanishes
continuously. Both the height and the width of the rarefaction
vanish, matching continuously the profile of the traditional
deflagration. In contrast, as the velocity approaches the Jouguet
detonation velocity $v_J^{\mathrm{det}}$, the shock wave becomes
thinner and \emph{higher}. Thus, the fluid velocity in front of the
wall is maximal for a Jouguet deflagration at
$v_w=v_J^{\mathrm{det}}$ (while the width of the shock wave
vanishes). On the contrary, for a detonation the fluid velocity
always vanishes in front of the wall. Therefore, the fluid velocity
is discontinuous as the solution changes from a Jouguet deflagration
to a detonation. This originates a jump in the parameters $\alpha_n$
and $\tilde{\eta}$ as functions of $v_w$, as can be seen in Figs.
\ref{figdetodefla} and \ref{figetaikkl}.

Leaving aside the jump of the parameters at $v_w=v_J^{\mathrm{det}}$,
one would expect that the supersonic Jouguet deflagration will always
fill the velocity gap between weak deflagrations and weak detonations
(as, e.g., in the left panels of Figs. \ref{figdetodefla} and
\ref{figetaikkl}). A fluid profile does exist for any value of the
wall velocity in the range $c\leq v_w\leq v_J^{\mathrm{det}}$. The
Jouguet deflagration reaches the value $v_w=c$ for the same values of
parameters as the weak deflagration, whereas the value
$v_w=v_J^{\mathrm{det}}$ is reached for a lower friction or a
stronger supercooling than the detonation. This behavior can be
explained by the fact that, for deflagrations, the compression wave
which propagates in front of the wall and reheats the fluid, causes a
friction effect \cite{ms09} which adds to the microphysics.
Technically, the friction effect for the deflagration arises as a
consequence of the relation (\ref{vfl}) between the nucleation
temperature $T_n$ and the reheated value $T_+$. For the detonation,
instead, we have $T_+=T_n$. This effect is easily estimated for weak
solutions \cite{ms09}. In the small supercooling limit (i.e., for
weak deflagrations) we have $v_w=\Delta p(T_n)/\eta_{\mathrm{eff}}$,
where $\Delta p$ is the pressure difference $p_--p_+$ and the
effective friction is given by $\eta _{\mathrm{eff}}=
({w_{-}}/{w_{+}})\left[ \eta +L^{2}/(\sqrt{3}w_{-})\right] $, which
doesn't vanish for $\eta=0$. In the ultrarelativistic limit, in
contrast, we have weak detonations with $v_w=1-\delta$, where
$\delta$ is proportional to $\eta^2$ and vanishes for vanishing
friction. For deflagrations with high wall velocities ($v_w\gtrsim
c$) the fluid velocity and temperature profiles have a sharp peak in
front of the wall, so the friction effect can be considerably large.
Notice that, in the right panel of Fig. \ref{figdetodefla}, the value
$v_J^{\mathrm{det}}$  is never reached by Jouguet deflagrations, even
in the limit $\tilde{\eta}\rightarrow 0$. Thus, we see that in some
cases the effective friction can prevent the supersonic deflagration
to reach the velocity of the Jouguet detonation.

Even in the cases in which the Jouguet deflagrations fill the whole
range between $c$ and $v_{J}^{\mathrm{det}}$,  only one of the
multiple solutions will be realized. As a consequence, there will
always be a gap in the velocity, as shown in Fig. \ref{figetaikkl}
(right panel). It is important to determine which of the possible
solutions will be realized in the phase transition as a stationary
state. In Refs. \cite{ikkl94,kl95,kl96}, the evolution from a given
initial configuration of the scalar field and the fluid variables
(i.e., from an initial ``bubble wall'') was studied by numerically
solving the time dependent partial differential equations. Thus,
their \emph{dynamical code} selected one of the possible final
states. It was observed that the strong deflagrations, as well as the
branch of detonations which are closer to the Jouguet point, are not
realized in the evolution of the bubble wall. Moreover, even if given
as initial conditions, these solutions transform into one of the
other solutions, suggesting an instability \cite{kl95}. Furthermore,
as seen in Figs. \ref{figdetodefla} and \ref{figetaikkl}, these
solutions have a non-physical behavior as functions of the parameters
\cite{ms09}. Therefore, they must be discarded. In contrast, weak
deflagrations, Jouguet deflagrations, and the branch of weaker
detonations are in general stable. For high friction or low
supercooling we can only have weak deflagrations.  As the parameters
are varied and the weak deflagration reaches the velocity $v_w=c$, we
must change to a Jouguet deflagration. On the other hand, for
$v_w\gtrsim c$ we will have to choose between a deflagration and a
detonation. Both are stable and both are reachable from different
initial conditions in the dynamical code of Ref. \cite{kl95}.
However, in the normal evolution of a bubble wall, the solution which
is realized is the detonation\footnote{Interestingly, the solutions
which are not realized are those which are closer to the speed of
sound and have a sharp peak in temperature. Thus, the dynamical
evolution selects the weaker stationary solution, i.e., the weak
detonation. For the fastest deflagrations that are realized, the
bubble wall first goes into a detonation configuration before
settling into a deflagration \cite{kl96}.}. Accordingly,  we must
always choose the detonation if possible; otherwise, the Jouguet
deflagration; otherwise, the weak deflagration. The result is shown
in the right panel of Fig. \ref{figetaikkl}, where we reproduced the
plot of the left panel, keeping only the selected solutions.

\begin{figure}[hbt]
\centering
\epsfysize=8cm \leavevmode \epsfbox{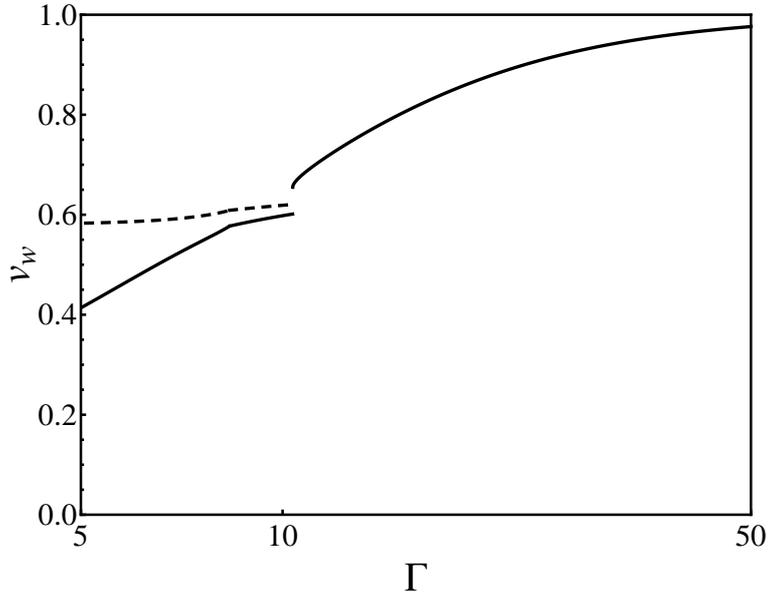}
\caption{The wall (solid) and shock (dashed) velocities as functions of
$\Gamma=1/\tilde{\eta}$. The values of the parameters are as in
Fig. \ref{figetaikkl}.}
\label{figgamma}
\end{figure}

Qualitatively, our results agree with the numerical solutions, even
for spherical bubbles (cf. Fig. 3 of Ref. \cite{kl96}). The
parameters we used correspond to Fig. 13 of Ref. \cite{ikkl94}.
There, the friction parameter is $\Gamma =1/ \tilde{\eta}$. For a
better comparison we plot the velocity as a function of $\Gamma$ in
Fig. \ref{figgamma}. For the range of friction considered in that
figure, we have a difference of at most a 5\% for weak deflagrations
and less for detonations. For the weakest solutions the difference
vanishes, as expected. Moreover, for the strongest solutions, i.e.,
those around the Jouguet point, the agreement is quite good, as the
jump in $ v_{w}$ is approximately at the same place, $\Gamma\approx
10$.

\begin{figure}[hbt]
\centering
\epsfysize=8cm \leavevmode \epsfbox{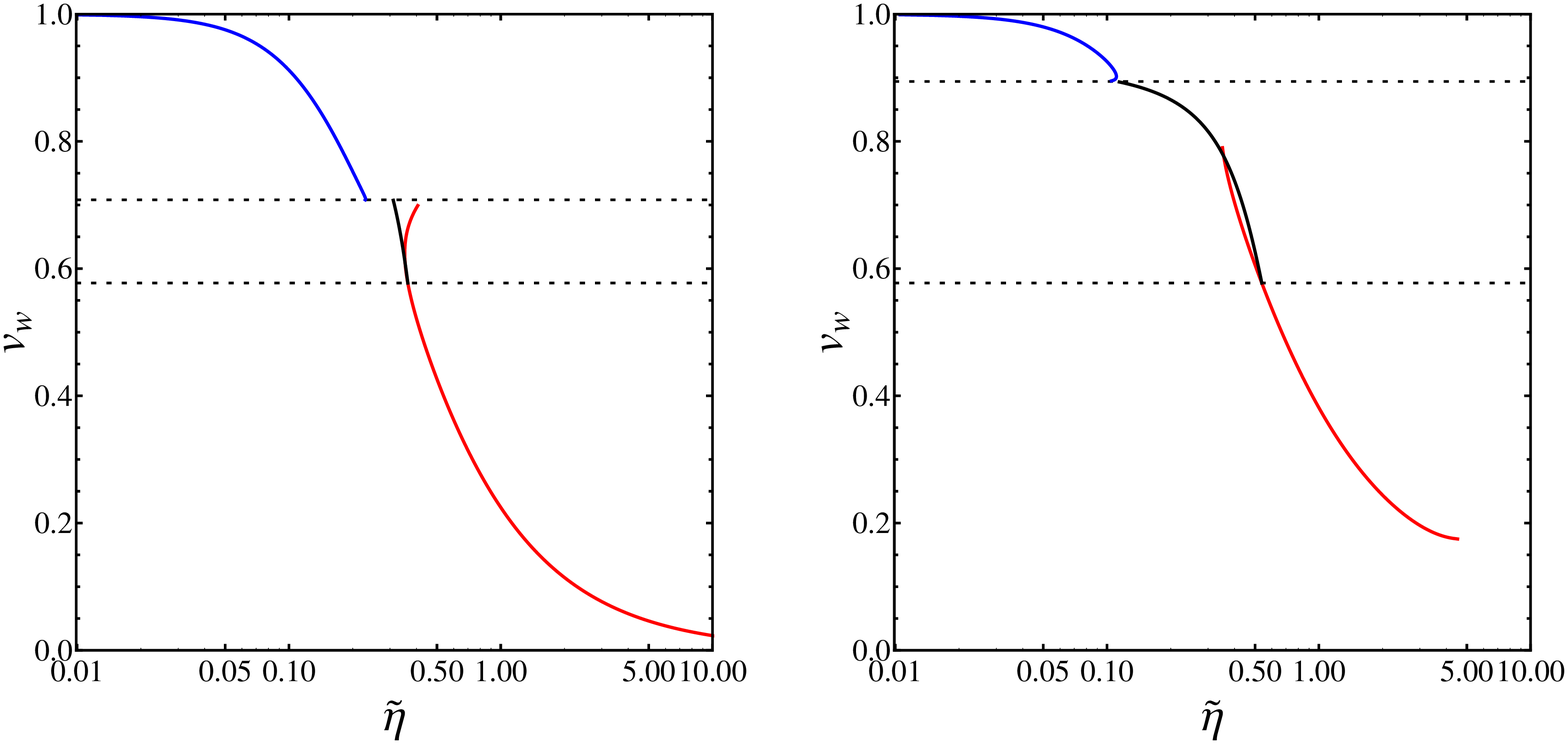}
\caption{The wall velocity  as a function of $\tilde{\eta}$
for $\alpha _{c}= 4.45\times 10^{-3}$ and $T_n/T_c=0.6$ (left panel), and
for $\alpha _{c}= 0.05$ ($a_-/a_+=0.85$) and $\alpha_n=0.5$ (right panel).
The dotted lines indicate
the speed of sound and the Jouguet detonation velocity.}
\label{fignofunca}
\end{figure}

One expects that for any set of values of the parameters there will
exist at least one solution. Consider a fixed value of $T_n$. For
large friction there should always be weak deflagrations and for
small friction there should always be weak detonations. In the
intermediate range there should exist Jouguet deflagrations. Although
this is in general true, we find some exceptions in extreme cases.
For instance, for a large amount of supercooling, $T_n/T_c=0.6$
(which is out of the range of Fig. \ref{figintersec}), we find that,
as we decrease the friction, the Jouguet deflagration velocity
reaches the value of the Jouguet detonation before the detonation
solution appears  (see Fig. \ref{fignofunca}, left panel). As a
consequence, there is a range of values of $\tilde{\eta}$ for which
there is no solution. For even stronger phase transitions, we find
that there may be no solutions for large values of the friction. This
behavior was observed also in Ref. \cite{ekns10}. This is shown in
the right panel of Fig. \ref{fignofunca}, where we considered values
of the parameters similar to those used in Ref. \cite{ekns10}. For
larger amounts of supercooling, there may be no subsonic
deflagrations at all, whereas supersonic deflagrations cease to exist
at some maximum friction.

We have checked that, for physical models, the parameters
(particularly the amount of supercooling) hardly fall into the case
of Fig. \ref{fignofunca}. In particular, we considered the
electroweak phase transition for several extensions of the Standard
Model \cite{lms12}. The situation of the right panel never arose, and
that of the left panel arose only in a few limiting cases which are
quite unlikely. For instance, extra scalars with very strong
couplings with the Higgs field may yield an exceedingly strongly
first-order phase transition. Only for some particular sets of
parameters, and for the highest values of the couplings of the extra
bosons, we found the situation of no stationary solution. Such
couplings are extreme in the sense that a little increase causes the
phase transition to remain stuck in the high-temperature phase and
the universe to enter a period of inflation. In spite of this, the
consequences of such models may be interesting and  deserve further
consideration. The absence of stationary solutions in some ranges of
parameters may be related to the existence of \emph{runaway}
solutions \cite{bm09}, which correspond to the wall propagating
ultra-relativistically, with the gamma factor $\gamma$ growing
linearly with time.  Next we discuss this possibility.

\section{Friction saturation and runaway solutions} \label{runaway}

The equation for the friction, Eq. (\ref{eqfield}), is obtained from
the equation of motion for the field $\phi$,
\begin{equation}
\partial _{\mu }\partial ^{\mu }\phi +\frac{\partial V}{\partial
\phi }+\sum_i \frac{dm^2_i}{d\phi}\int\frac{d^3p}{(2\pi)^32E_i}f_i(p)=0,
\label{eqmicro}
\end{equation}
where $V$ is the zero-temperature effective potential and $f_i$ is
the distribution function of particle species $i$. The latter can be
decomposed into the equilibrium distribution function
$f_i^{\mathrm{eq}}$ and a deviation $\delta f_i$. The equilibrium
part of $f_i$ in Eq. (\ref{eqmicro}) gives $\partial
V_T/\partial\phi$, where $V_T$ is the thermal part of the
finite-temperature effective potential (i.e., $\mathcal{F}=V+V_T$).
Together with the term $\partial V/\partial\phi$ this gives $\partial
\mathcal{F}/\partial\phi$, which is the second term in Eq.
(\ref{eqfield}), whereas the deviation gives the friction term.
Usually, the deviations from equilibrium are assumed to be small,
corresponding to a non-relativistically moving wall. This gives a
friction force proportional to the velocity of the wall with respect
to the plasma. Such microphysical calculations can be used to
determine the value of the coefficient $\tilde{\eta}$ and the
function $f(\phi/T)$ in Eqs. (\ref{eqfield}) and  (\ref{contfric}).
In this section we shall set for simplicity $f(\phi/T)=1$. Therefore,
this procedure gives a friction force per unit surface area
\begin{equation}
\frac{F_{\mathrm{fr}}}{A}=\tilde{\eta}T\sigma v\equiv\eta v,
\label{fricnr}
\end{equation}
as explained at the end of section \ref{hydro}. Here, $v$ is the
(negative) fluid velocity in the wall frame. This corresponds to a
friction term $\tilde{\eta}Tv\partial_x\phi$ in the field equation
(\ref{contfric}). The simplest relativistic generalization is the
$u^{\mu}\partial_{\mu}\phi$ term in Eq. (\ref{eqfield}).

Recently, the opposite limit was considered in Ref. \cite{bm09}. The
friction acting on the electroweak bubble wall was derived for a wall
which is already propagating ultra-relativistically, with gamma
factor $\gamma\sim10^9$. Such a fast moving wall validates a number
of approximations. The reflection coefficients are exponentially
suppressed. In the frame of the wall, incoming particles have
received no signal that the wall is approaching and are in
equilibrium. Furthermore, interactions or scatterings between plasma
particles can be neglected and the occupancies evolve undisturbed.
Therefore, only the equilibrium occupancies in the symmetric phase
are used in the calculation. The occupancies are assumed to be
constant along a classical trajectory through the bubble wall.

As a consequence, the resulting force on the wall does not have a
velocity-dependent term. The net force per unit area on the wall is
found to be given by
\begin{equation}
\frac{F}{A}=V(\phi_+)-V(\phi_-)-\sum_i[m^2_i(\phi_-)-m^2_i(\phi_+)]
\int\frac{d^3p}{(2\pi)^32E_{i+}}f^{\mathrm{eq}}_{i+}(p).
\end{equation}
This means that the total force per unit area is given by the
``pressure difference''
\begin{equation}
\tilde{p}_--\tilde{p}_+=-\tilde{\mathcal{F}}_-+\tilde{\mathcal{F}}_+,
\label{totalforce}
\end{equation}
where $\tilde{\mathcal{F}}(\phi,T)$ is the mean field approximation
to the effective potential, which is obtained by keeping only the
quadratic terms in a Taylor expansion of $V_T$ about $\phi_+$
\cite{bm09,ekns10}
\begin{equation}
\tilde{\mathcal{F}}(\phi,T)=V(\phi)+V_T(\phi_+)+\sum_i
[m^2_i(\phi )-m^2_i(\phi_+)] \left. \frac{dV_T}{dm^2_i}\right|_{\phi_+}.
\end{equation}
To determine whether or not the wall can run away,
$\tilde{\mathcal{F}}$ must be used instead of $\mathcal{F}$, and the
total force (\ref{totalforce}) must be positive, i.e., \emph{if
replacing $V_T(\phi)$ with its second-order Taylor approximation
removes the minimum $\phi_-$  or raises it above the minimum
$\phi_+$, the bubble wall cannot run away} \cite{bm09}.

In particular, in Ref. \cite{bm09} it is shown that the bubble wall
never runs away in a ``fluctuation induced'' first-order phase
transition, i.e., a phase transition which is first-order due to the
thermal part of the potential (e.g. the MSSM). As a simple example,
consider the high-temperature expansion
\begin{equation}
V_T(\phi)=\sum_i\frac{T^2m^2_i(\phi)}{24}-\frac{Tm_i^3(\phi)}{12\pi}
+\mathcal{O}(m^4). \label{vpot}
\end{equation}
It is well known that the cubic term in (\ref{vpot}) may cause a
first-order phase transition. This term is not present in the mean
field potential. If the first-order character of the phase transition
is due only to this term, then in the mean field potential the
``broken symmetry'' minimum $\phi_-\neq 0$ raises above the
``symmetric'' minimum $\phi_+=0$. In such a model the wall will reach
a terminal velocity $v_w<1$. An example of a model which does not
rely on the terms $Tm_i^3(\phi)$ to yield a first-order phase
transition is a potential with tree-level cubic terms. This is
possible, e.g. in extensions of the Standard Model with singlet
scalar fields, as considered in Ref. \cite{bm09}.

In this regime the friction force does not depend on the velocity of
the fluid relative to the wall, whereas for $v\to 0$ it is
proportional to $v$. It would be important for applications to know
the behavior of the friction for intermediate velocities.
Unfortunately, this constitutes a nontrivial problem. A simple
interpolation between the two regimes was considered in Ref.
\cite{ekns10}. The approximation consisted in the replacement
$u^{\mu}\partial_{\mu}\phi\;\to\;{u^{\mu}\partial_{\mu}\phi}\,/{\sqrt{1+
(\lambda_{\mu}u^{\mu})^2}} $  in Eq. (\ref{eqfield}), with
$\lambda_{\mu}=(0,0,0,1)$ in the wall frame. This is equivalent to
the replacement $v\gamma\, \partial_{x}\phi\to v\, \partial_{x}\phi$.
This modification does not alter the discussion of section
\ref{hydro}. Furthermore, it simplifies the analytic equations of
section \ref{bag} through the replacement $ \left\vert
v_{+}\right\vert \gamma _{+}+\left\vert v_{-}\right\vert \gamma
_{-}\to \left\vert v_{+}\right\vert+\left\vert v_{-}\right\vert$ in
Eq. (\ref{fricbag}).

\begin{figure}[hbt]
\centering
\epsfysize=8cm \leavevmode \epsfbox{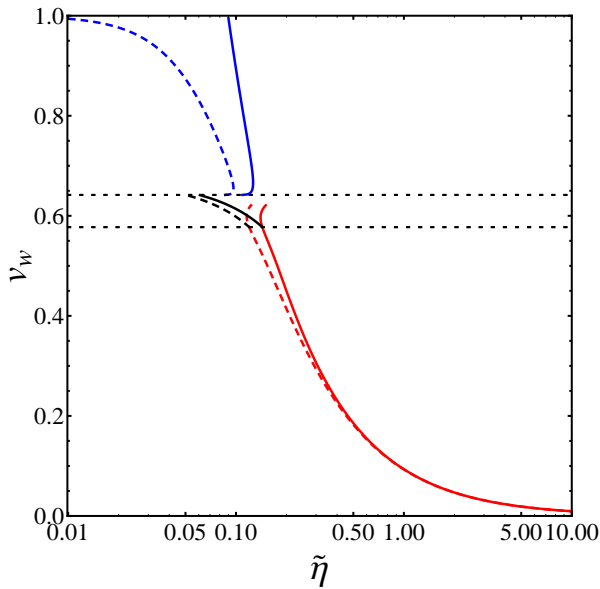}
\caption{The wall velocity for the parameters of Fig. \ref{figetaikkl},
with a friction force given by $\eta v$ (solid) and $\eta \gamma v$ (dashed).
The dashed line is the curve of Fig. \ref{figetaikkl}.}
\label{figekns}
\end{figure}

This phenomenological approach for the friction changes significantly
the behavior in the detonation regime. As an example,  we show  in
Fig. \ref{figekns} how the results of Fig. \ref{figetaikkl} are
modified.
Notice that detonations exist only in a small interval of the
friction parameter. For higher values of the friction we have
deflagrations, whereas for smaller values there is no stationary
solution and  the bubble wall runs away. These results are in
accordance with those of Ref. \cite{ekns10}. In particular, we obtain
detonations only in a narrow region in the $\eta \alpha_n$-plane.

\begin{figure}[hbt]
\centering
\epsfysize=8cm \leavevmode \epsfbox{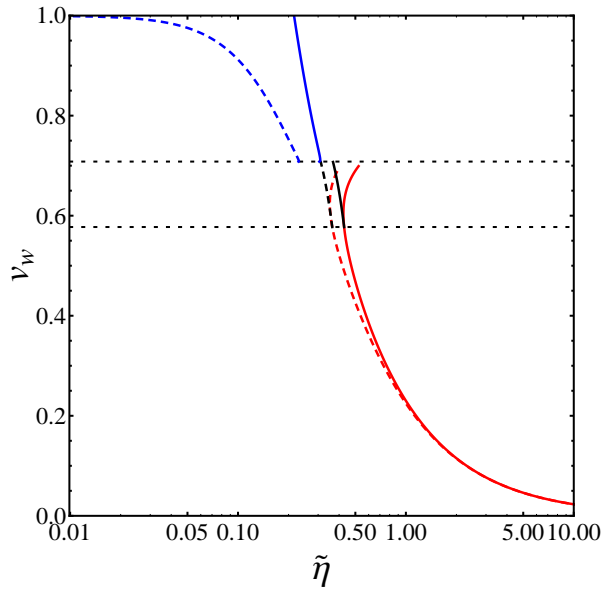}
\caption{The wall velocity  as a function of $\tilde{\eta}$
for $\alpha _{c}= 4.45\times 10^{-3}$ and $T_n/T_c=0.6$,
with a friction force given by $\eta v$ (solid) and $\eta \gamma v$ (dashed).}
\label{fignofunca2}
\end{figure}

On the other hand,  the modification of the friction term does not
affect significantly the deflagration solutions, as expected. In the
previous section we have found regions of parameters corresponding to
deflagrations, where there is no stationary solution (see Fig.
\ref{fignofunca}). In particular, the situation of the left panel of
Fig. \ref{fignofunca} may arise in some very strongly first-order
phase transitions. With the modified friction, this behavior remains
(see Fig. \ref{fignofunca2}). This is due to the fact that the
deflagration solutions are not significantly altered by the
modification of the friction. The runaway solutions appear instead
for smaller values of the friction, corresponding to detonations.
Nevertheless, the no-solution region has shrunk with this
approximation, and it is possible that for a better approximation it
will not exist at all.

We wish to stress that this approximation, which consists essentially
in assuming a friction force of the form (\ref{fricnr}) for any wall
velocity, may be too simplistic. Although it reproduces the
saturation of the friction force at large $\gamma v$, the friction
force saturates too soon (at values $\gamma\sim 1$) to its
ultra-relativistic value (which is in principle valid for $\gamma\sim
10^9$). A different variation of the friction may allow the existence
of stationary solutions (i.e., detonations) for a wider range of
parameters. In particular, it is not clear that the friction should
be a monotonically growing function of $v$. Indeed, the assumptions
that lead to the runaway solution are based on the fact that the
fluid is rather unaffected by the passage of the wall, due to the
high speed of the latter. Conversely, the fact that the wall is
unaffected by the fluid allows it to accelerate. This is somewhat
similar to what happens at the macroscopic level with the stationary
hydrodynamical modes, namely, the fastest the detonation, the weaker
the disturbance it causes on the fluid \cite{ekns10,lm11}. Moreover,
solutions with intermediate velocities (around the Jouguet point)
cause the maximum disturbance. Weaker deflagrations have small
velocities and cause little perturbations. Weaker detonations cause
less perturbations of the fluid and can thus move much faster.  We
may expect a similar behavior at the microscopic level, i.e., that
intermediate velocities will cause larger departures from the
equilibrium distributions and, thus, a higher friction.

In particular, it may happen that the real friction force is well
approximated by $\eta \gamma v$ up to relatively high values of
$\gamma v$, such that $\gamma \gg 1$ but not yet as large as to
fulfill the hypothesis of Ref. \cite{bm09}. If this is the case, then
the wall will end up moving with a terminal velocity with a moderate
value of $\gamma$ and never reach the ultra-relativistic regime, in
spite of the existence of runaway solutions. In such a case, the
model $F_{\mathrm{fr}}/A=\eta \gamma v$ considered in the previous
section would give a better approximation than $\eta v$, although the
latter gives the correct ultra-relativistic behavior. Furthermore, it
seems that this approximation does not take into account the fact
that, if the first-order nature of the phase transition is
fluctuation induced, then the bubble wall should not run away, even
if the friction coefficient is very small. In such a case we expect
the dashed line in Fig. \ref{figekns} to give the correct behavior.

\section{Conclusions} \label{conclu}

We have investigated the steady state motion of phase-transition
fronts in a cosmological first-order phase transition. Our main goal
was to find analytical approximations for the wall velocity, taking
into account the different possibilities for the hydrodynamical modes
and fluid profiles. Therefore, we have considered the case of planar
walls, which allow to obtain analytical approximations. In Ref.
\cite{ms09} we considered the cases of weak deflagrations preceded by
a shock front and weak detonations followed by a rarefaction wave.
Here, we have studied also the case of Jouguet deflagrations which
have both shock and rarefaction waves and move supersonically. We
have considered two different phenomenological models for the
friction. One of them grows linearly with the relativistic velocity
$\gamma v$ \cite{ikkl94}, and the other saturates for large $\gamma
v$ \cite{ekns10}. The latter reproduces the behavior of the friction
force in the ultra-relativistic limit and leads to runaway solutions
\cite{bm09}.

Our main result is a set of  algebraic equations which allow to
obtain, from the thermodynamic parameters and the friction
coefficient, the value of the wall velocity which will be realized as
the final stationary state. These analytical results rely on several
approximations, such as the use of the bag equation of state and the
thin wall approximation. Implementing the latter in the equation for
the friction requires some ansatz for the variation of the entropy
density inside the wall and also for that of the fluid velocity. The
approximation for the entropy density seems to be the roughest one,
as we do not obtain the curve of zero entropy production in the limit
of vanishing friction. By comparing with numerical lattice
calculations \cite{ikkl94,kl95,kl96}, we have checked that the
strongest departure from the exact solution occurs for the strongest
physical solutions, i.e., those around the Jouguet points (either for
detonations or deflagrations).

For a friction of the form $\gamma v$, our results are in good
quantitative agreement with the cases of planar walls considered in
Refs. \cite{ikkl94,kl95} and in good qualitative agreement with those
of spherical bubbles considered in Refs. \cite{kl95,kl96}. For a
friction which saturates for large $\gamma$, the wall velocity shows
essentially the same behavior as in Ref. \cite{ekns10}, which
considered spherical bubbles. We remark that this latter
approximation, although reproducing the correct behaviors for $v\to
0$ and for $v\to 1$, may still be too simplistic for intermediate
velocities. In particular, the friction saturates to a constant value
for relatively small velocities, i.e., for values of the gamma factor
which are much smaller than those which justify the approximations
that lead to the runaway solution \cite{bm09}. As a consequence, the
region of parameter space in which the bubble wall runs away may be
largely overestimated. The problem of determining the behavior of the
friction at intermediate velocities is a difficult one and deserves
further investigation, since it has important implications for
cosmology.

\section*{Acknowledgements}

This work was supported in part by Universidad Nacional de Mar del
Plata, Argentina, grants  EXA 473/10 and 505/10. The work by A.M. was
supported by CONICET through project PIP 112-200801-00943. The work
by A.D.S. was supported by CONICET through project PIP
122-201009-00315.

\end{document}